\documentclass[preprint]{aastex}

\shorttitle{Point-Source Filter}
\shortauthors{Perera et al.}

\usepackage{graphicx}
\usepackage{amsmath}

\begin{document}

\title{An Efficient and Optimal Filter for Identifying Point Sources
  in Millimeter/Sub-Millimeter Wavelength Sky Maps}

\author{T.~A. Perera}
\email{tperera@iwu.edu}
\author{J.~R. Schaar}
\author{A. Mancera}
\affil{Department of Physics, Illinois Wesleyan University, Bloomington, IL 61702}

\author{G.~W. Wilson}
\affil{Department of Astronomy, University of Massachusetts Amherst,
  Amherst, MA 01003}

\author{K.~S. Scott}
\affil{National Radio Astronomy Observatory, Charlottesville, VA 22903}

\author{J.~E. Austermann}
\affil{Center for Astrophysics and Space Astronomy, Boulder, CO 80309}

\begin{abstract}
A new technique for reliably identifying point sources in
millimeter/sub-millimeter wavelength maps is presented.  This method
accounts for the frequency dependence of noise in the Fourier domain
as well as non-uniformities in the coverage of a field.  This optimal
filter is an improvement over commonly-used matched filters that
ignore coverage gradients.  Treating noise variations in the Fourier
domain as well as map space is traditionally viewed as a
computationally intensive problem.  We show that the penalty incurred
in terms of computing time is quite small due to casting many of the
calculations in terms of FFTs and exploiting the absence of sharp
features in the noise spectra of observations.  Practical aspects of
implementing the optimal filter are presented in the context of data
from the AzTEC bolometer camera.  The advantages of using the new
filter over the standard matched filter are also addressed in terms of
a typical AzTEC map.
\end{abstract}


\section{Introduction}

The discovery and study of sub-millimeter galaxies, or SMGs, has
become a key enterprise within millimeter/sub-millimeter (mm/sub-mm)
astronomy over the past decade or so \citep[see review
  by][]{blain2002}. With bolometric luminosities $\gtrsim
5\times10^{12}\,L_\odot$ and star formation rates $\gtrsim
100\,M_\odot$\,yr$^{-1}$ , SMGs represent some of the most luminous
galaxies in the early Universe. The strong negative $k$-correction at
$\lambda > 500\,\mu$m means that a galaxy of a given luminosity will
be equally detectable in a flux-limited survey from $1 \lesssim z
\lesssim 10$. A deep, wide-area survey at mm/sub-mm wavelengths is
thus a sensitive probe of starburst galaxies from the epoch at which
they first turn on through the peak of star formation activity in the
Universe at $z \approx 1-2$. Several papers have reported on the
source counts of SMGs detected at $\lambda = 250 - 2000\,\mu$m
\citep[e.g.][and references
  therein]{coppin2006,Bertoldi2007,weiss2009,vieira2010,scott2012},
which provide strong constraints on the evolutionary history of
massive galaxies
\citep[e.g.][]{baugh2005,valiante2009,bethermin2012}. SMG surveys from
single-dish telescopes also provide a catalog of interesting targets
to follow-up with higher angular resolution imaging of the dust and
molecular gas in these objects, which inform on the physical processes
that trigger and maintain the starbursts in these galaxies.

An important criterion for the success of such surveys is the design
of a data analysis scheme that can reliably identify SMG candidates on
sky maps. It is especially important to minimize the rate of false
detections because they can result in a waste of valuable time and
resources in follow-up observations. But reliably identifying
astronomical signal in mm/sub-mm maps is a difficult task because
these maps are inherently low in signal/noise due to atmospheric
contamination, instrument noise, and the high confusion limit of
surveys from typical mm/sub-mm telescopes ($\sim$10\,m diameter). In
the wide field surveys designed to detect SMGs, they usually appear as
point sources much smaller than the angular resolution of the
telescopes used. A simple approach to identify point sources that is
used by standard data analysis packages \citep[see][for
  example]{Stetson1987} is: 1) smooth the signal map by convolving it
with the point spread function (PSF); 2) assume that the errors
associated with pixels are uncorrelated and propagate them through the
convolution; and 3) select the peaks in the smoothed map based on the
per-pixel signal/noise determined from steps 1) and 2). This reduces
high-frequency signal variations between adjacent pixels and thus
increases the signal/noise for point-source detection.

However, there are two general problems with real mm/sub-mm maps that
renders the above procedure inadequate for reliably identifying point
sources. First, the noise in these maps is generally not white; it is
more pronounced on larger scales or at low (spatial) frequencies. This
effect, which gives rise to pixel-pixel correlations, is due mainly to
$1/f$ drifts in atmospheric and/or instrumental conditions. In
mm/sub-mm astronomy, this problem is usually treated with a
``matched filter'' implemented in the Fourier domain
\citep[e.g.][]{Tegmark1998, Barreiro2003, Vio2004, Barnard2004,
  Perera2008, Chapin2011}. However, this filtering technique is
optimal only in the case of {\em uniform} coverage, which brings up
the second problem with real mm/sub-mm maps: the coverage of a field
is non-uniform in general.  Because these observations are usually
carried out by scanning an array of detectors across the field of
interest, variations in atmospheric conditions or detector noise
during the scan are often responsible for this non-uniformity.  With
commonly used schemes such as raster- or Lissajous-scanning, coverage
also tends to decrease smoothly from the center to the edges of the
map.

The correct way to deal with both these problems involves the
construction and inversion of a pixel-pixel noise covariance matrix.
This path can be extremely challenging computationally for maps
containing $\sim 100,000$ pixels.  Therefore, a common strategy is to
pick out a region of the map that has essentially uniform coverage and
then apply the standard matched filter to it.  Even when a
near-uniform region exists, ignoring small coverage variations within
it can have a noticeable effect, as we will show below.  Furthermore,
despite gradients, the coverage may be deep enough near the edges of
the map to identify bright sources with high significance, and these
``border'' regions can often cover a significant area compared to the
near-uniform region.  In some cases, significant coverage gradients
are unavoidable due to difficult observing conditions or an observing
strategy where a mosaic of small maps are stitched together \citep[for
  example, see][]{Borys2003, coppin2006, Austermann2010}.  A truly
optimal analysis would help improve the reliability with which
point-source candidates are identified in {\it all} regions of a map
despite coverage gradients.

In this paper, we present an algorithm that addresses non-uniform
coverage as well as low-frequency noise in a computationally efficient
manner.  This technique is designed to be optimal in the regime where
blending of resolved sources is negligible.  The work presented here
is an extension of the standard data reduction tools used for
point-source extraction in maps taken with the 1.1\,mm bolometer array
camera, AzTEC \citep{Wilson2008}. While this builds on the existing
AzTEC reduction pipeline \citep{Scott2008, Perera2008}, the principles
are generic to any scan-data from mm/sub-mm arrays. In section
\ref{outline} we present the formal methodology that will be used to
identify point sources and to estimate their brightness. Essentially,
this method involves a least-squares fit of the PSF to every point on
the map. In section \ref{simple_filter}, we demonstrate that, for the
case of uniform coverage, the fit paradigm converges to the familiar
and computationally efficient matched filter routinely used in
point-source searches.  In section \ref{general_filter}, we develop an
implementation of the pixel-by-pixel fit that can efficiently handle
the case of non-uniform coverage. Then, in section \ref{application},
we discuss the practicalities of applying this technique to AzTEC
data.  Finally, in section \ref{conclusion}, we conclude by discussing
the performance and scope of the optimal filter introduced here.

\section{The Method}
\label{outline}

We leverage our knowledge of the shape of isolated point sources in
our maps to construct an {\em optimal} or {\em wiener} filter using
the point spread function (PSF) as the template.  Applying this filter
to the map is formally equivalent to centering the PSF on each pixel
of the map and fitting for the best-fit amplitude \citep{Stetson1987}.
In our case, we store the best fit amplitude at each pixel in a
separate map which we will refer to as the {\em filtered signal map.}
In addition, we construct a new map of the same dimensions as the
signal map which contains 1/error$^2$ estimates of the best fit
amplitudes.  This map will be referred to as the {\em filtered
  coverage map.}  These maps, augmented by noise realizations of the
field, are the primary inputs to the identification of point sources.

\subsection{AzTEC data as an example}

AzTEC is a 144-element semiconductor-type bolometer array that imaged
the sky at 1.1-mm wavelength over two successful observing campaigns:
one using the 15-m James Clark Maxwell Telescope (JCMT) in Hawaii from
2005-2006 \citep{Wilson2008}, and the other using the 10-m Atacama
Sub-millimeter Telescope Experiment (ASTE) in the Atacama desert of
Chile from 2007-2008 \citep{Ezawa2008}. Over these two observing runs,
close to a hundred fields were mapped, and each map comprises 100,000
pixels or more.  Therefore, computational speed was an important
consideration when developing this algorithm.  Computational speed
will be even more important for newer and upcoming mm/sub-mm
observatories like the Large Millimeter Telescope and CCAT that can
image the sky at even faster rates with higher angular resolution.

\begin{figure}[h!]
\centering
\leavevmode
\begin{tabular}{cc}
\includegraphics[height=2.9in]{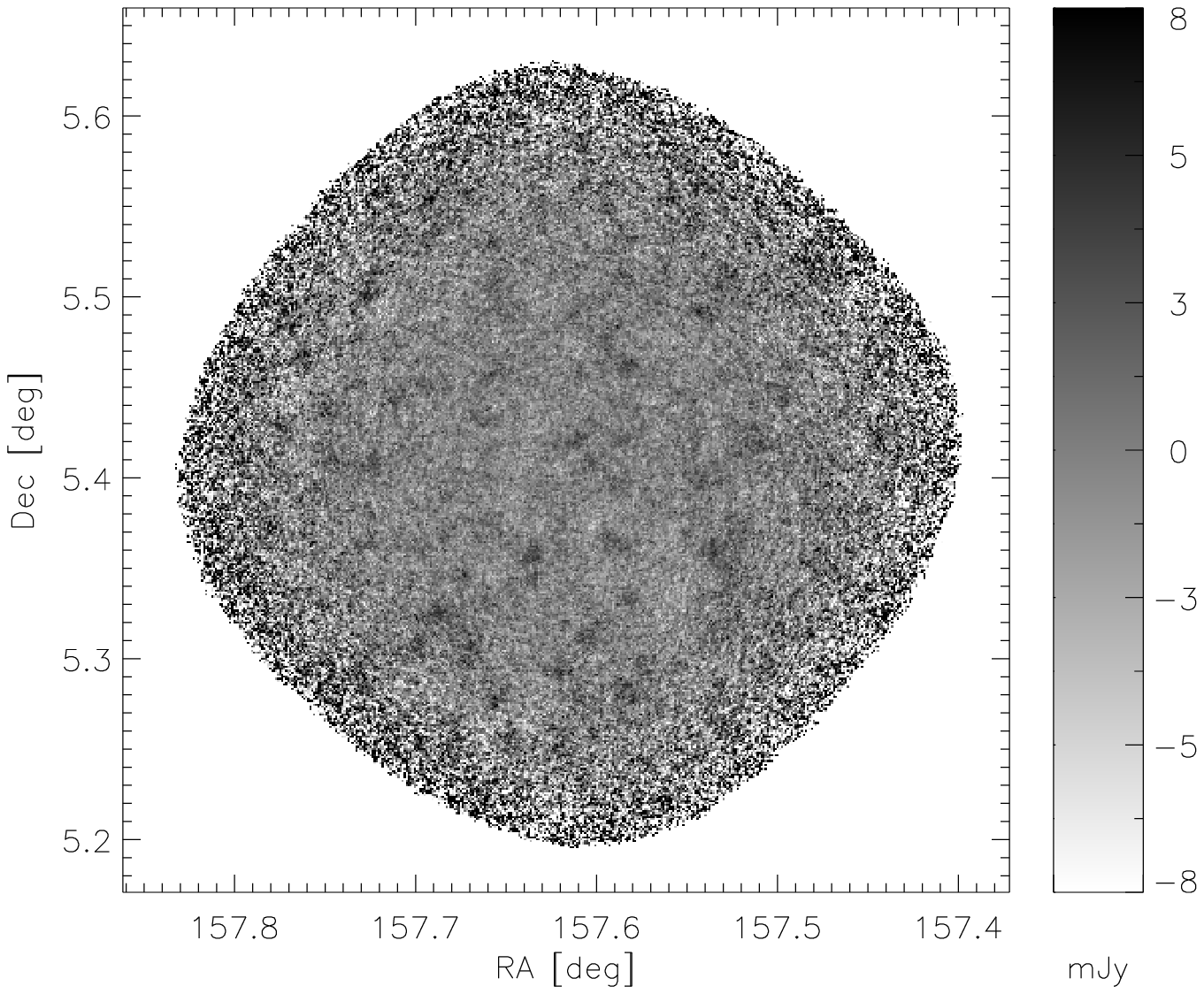}
& 
\includegraphics[height=2.9in]{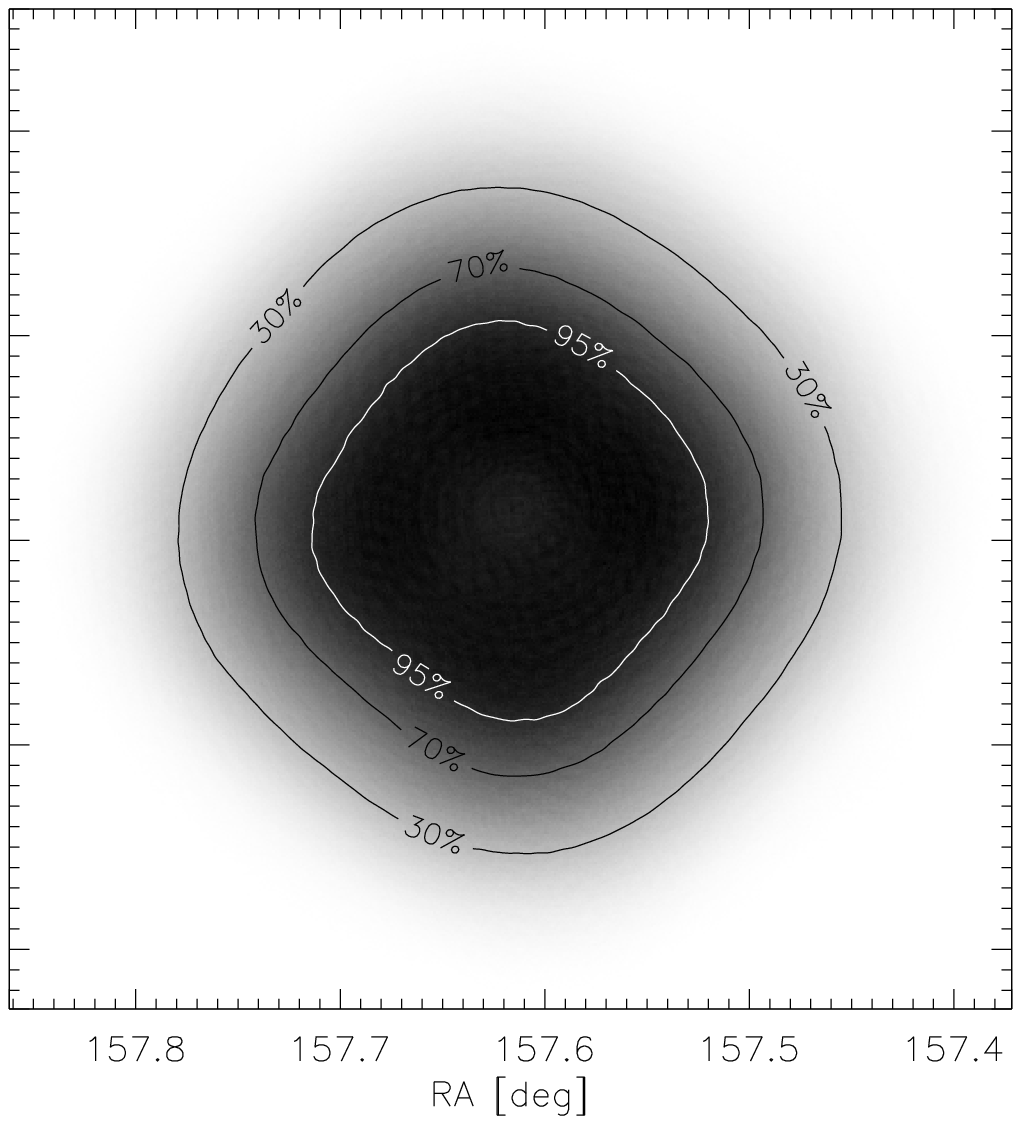}\\
(a) & (b)
\end{tabular}
\caption{(a) 1.1mm-wavelength map of a region surrounding the quasar
  SDSSJ1030+0524, obtained with AzTEC-ASTE.  The outer regions are
  saturated high (black) or low (white) on an astronomically
  interesting scale ($\pm8\,$mJy).  Astronomical features are visible
  in the central unsaturated (grey scale) region, mainly as PSF-sized
  dark patches (``blobs''), or point sources. (b) Coverage map of this
  field.  It is filled with the values $1/\epsilon^2({\bf x_p}),$
  where $\epsilon({\bf x_p})$ is the uncertainty of the signal at the
  $p$'th pixel of (a).  Coverage contours are labeled relative to the
  maximally covered (dark) central regions.
\label{raw_maps}
}
\end{figure}
To illustrate the starting point for the filtering process, we present
in Fig.~\ref{raw_maps} an unfiltered AzTEC signal map and the
corresponding coverage map.  In the AzTEC data analysis, coverage maps
have units of 1/Jy$^2$ and are filled with estimates of 1/error$^2$ at
each pixel.  In Fig.~\ref{raw_maps} and throughout the rest of this
paper, we will use maps and plots related the AzTEC-ASTE data on a
region centered around the high-redshift quasar SDSSJ1030+0524 to
illustrate our points. This data set consists of 45 observations of
the field, each lasting about forty minutes, carried out over November
2008.  Scientific results from this AzTEC-ASTE field are published in
\citet{Humphrey2011} and \citet{Zeballos2013}.  These data were taken
by scanning the AzTEC array in a Lissajous pattern, and results in a
map with deepest coverage in the center, and decreasing coverage
toward the edges of the map.

\subsection{Assumptions}
\label{outline:assumptions}

Two assumptions, both of which are generally applicable to most
observations, are critical for the efficient implementation of the
process described below.  They are:

\noindent
1. {\em The effective PSF does not change with location on the map.}
We represent the PSF by the two-dimensional function $f({\bf x})$.
The value of this function at the $i$'th pixel when the PSF is
centered on the $p$'th pixel will be denoted by $f({\bf x_i} - {\bf
  x_p}),$ where ${\bf x_i}$ and ${\bf x_p}$ are 2-d vectors that
specify the locations of the $i$'th and $p$'th pixels, respectively.
We note that while this assumption is explicitly not valid for some
instruments (e.g., the {\em Chandra} X-ray telescope), this is a
reasonable assumption for many imaging instruments.

\noindent
2. {\em Noise-induced correlations between two pixels, hereafter
  denoted $Corr({\bf x_k}, {\bf x_l}),$ depend only on the distance
  $|{\bf x_k} - {\bf x_l}|,$ between the pixels.}  That is, neither
the location of the two pixels on the map nor their relative
orientation determines the correlation in their noise.  In such cases,
$Corr({\bf x_k}, {\bf x_l})$ may be conveniently expressed as the
inverse Fourier transform (IFT) of the {\em power spectral density}
(PSD) of noise in the map.  If we denote the PSD by $V^2({\bf k_a}),$
where ${\bf k_a}$ is a 2-d vector in the Fourier domain of the map,
\begin{equation}
Corr({\bf x_k}, {\bf x_l}) = \sum_{a}^{N_\mathrm{pixel}}
V^2({\bf k_a})~\mathrm{exp} [ 2 \pi j {\bf k_a} \cdot ( {\bf x_k} - {\bf
    x_l} ) ],
\label{Vsq}
\end{equation}
where $N_\mathrm{pixel}$ is the total number of pixels in the map, and
the sum is over all $N_\mathrm{pixel}$ vectors ${\bf k_a}$.  (Appendix
A contains a brief description of the discrete Fourier transform
conventions and their corollaries used here.)

In Eq.~\ref{Vsq}, $V^2({\bf k_a})$ is a normalized form of the PSD
that satisfies $\sum_{a}^{N_\mathrm{pixel}} V^2({\bf k_a}) = 1,$ so
that the diagonal elements of the correlation matrix evaluate to 1.
Note that for a completely flat PSD ($V^2({\bf k_a})={\rm cnst}$) the
correlation matrix is diagonal.  In most observations the noise PSD
varies smoothly with ${\bf k}$ and has {\em broad} features rather than
narrowly peaked ones (see Fig.~\ref{psd}, for example).  In such cases
the correlation matrix is band-diagonal (i.e., $Corr({\bf x_k}, {\bf
  x_l}) \simeq 0$ when $|{\bf x_k} - {\bf x_l}|$ is large).  This is a
property that we will exploit below in our pursuit of computational
efficiency.

\subsection{An optimal filter via a generalized least-squares fit}

We implement the optimal filter by minimizing the quantity
\begin{equation}
\chi_p^2 = \sum_{k,l=0}^{N_\mathrm{pixel}}[d({\bf x_k}) - s_p f( {\bf
    x_k} - {\bf x_p} )] W_{kl} [ d({\bf x_l}) - s_p f({\bf x_l} - {\bf
    x_p}) ],
\label{chisq1}
\end{equation}
for each pixel, $p$, in the map where $d({\bf x_k})$ and $d({\bf
  x_l})$ denote the value of the unfiltered signal map at the $k$'th
and $l$'th pixels, $s_p$ is the amplitude of the fitted PSF, and $W$
is an $N_\mathrm{pixel} \times N_\mathrm{pixel}$ {\em weight} matrix
whose calculation is described below.  The summations in
Eq.~\ref{chisq1} are over all pixels of the map.  Eq.~\ref{chisq1} is
minimized when
\begin{equation}
s_p = \frac{\displaystyle{ \sum_{k,l} W_{kl} d({\bf x_k}) f({\bf x_l -
      {\bf x_p}}) }} {\displaystyle{ \sum_{k,l} W_{kl} f({\bf x_k -
      {\bf x_p}}) f({\bf x_l - {\bf x_p}}) }}.
\label{s_p}
\end{equation}

For least-squares fits, the minimum variance choice of the weight
matrix is $W=C^{-1}$ where $C$ is the pixel-pixel noise
covariance matrix (i.e. the map's covariance in the absence of
signal).  In this case, the $\chi_p^2$ calculated in Eq.~\ref{chisq1}
is drawn from a true $\chi^2$ distribution with $N_{\rm pixels}-1$ degrees
of freedom.  The elements of $C$ are
\begin{equation}
C_{kl} = \epsilon({\bf x_k}) Corr({\bf x_k}, {\bf x_l}) \epsilon({\bf
  x_l}) = \displaystyle{ \epsilon({\bf x_k}) \left( \sum_{a} V^2({\bf
    k_a}) e^{ 2 \pi j {\bf k_a} \cdot ( {\bf x_k} - {\bf x_l} ) }.
  \right) \epsilon({\bf x_l}) }
\label{Ckl}
\end{equation}
where $\epsilon({\bf x_k})$ and $\epsilon({\bf x_l})$ are the standard
deviations of noise in the $k$-th and $l$-th pixels respectively.
Note that the diagonal elements, $C_{kk}$ evaluate to $\epsilon^2({\bf
  x_k}),$ as they should.

The elements of the weight matrix can then be calculated as
\begin{equation}
\displaystyle{
W_{kl} = [C^{-1}]_{kl} = {1 \over N_\mathrm{pixel}^2}
{1 \over \epsilon({\bf x_k})} \left(
\sum_{a} \frac
{e^ { 2 \pi j {\bf k_a} \cdot ( {\bf x_k} - {\bf x_l} ) }}
{V^2(\bf k_a)}
\right) {1 \over \epsilon({\bf x_l})}
}.
\label{Wkl}
\end{equation}
Using Eq.~\ref{app_conv} of Appendix A, it is straightforward to
verify that $\sum_l W_{kl} C_{lm} = \delta_{km}.$  Substituting the
above form of $W_{kl}$ into Eq.~\ref{s_p}, we may express the numerator
and denominator of Eq.~\ref{s_p} as
\begin{eqnarray}
N_p & = & {1 \over N_\mathrm{pixel}^2}
\sum_{k,l}
{f({\bf x_k} - {\bf x_p}) \over \epsilon({\bf x_k})} 
\left(
\sum_{a}
\frac{e^ { 2 \pi j {\bf k_a} \cdot ( {\bf x_k} - {\bf x_l} ) }}
{V^2(\bf k_a)}
\right)
{d({\bf x_l}) \over \epsilon({\bf x_l})}
\label{Ngen}
\\
D_p & = & {1 \over N_\mathrm{pixel}^2}
\sum_{k,l}
{f({\bf x_k} - {\bf x_p}) \over \epsilon({\bf x_k})} 
\left(
\sum_{a}
\frac{e^ { 2 \pi j {\bf k_a} \cdot ( {\bf x_k} - {\bf x_l} ) }}
{V^2(\bf k_a)}
\right)
{f({\bf x_l} - {\bf x_p}) \over \epsilon({\bf x_l})}.
\label{Dgen}
\end{eqnarray}
In practice, we form separate ``maps'' corresponding to $N$ and
$D$ because the latter is useful by itself (see below).  In doing so,
the nested sums of Eq.~\ref{Ngen} and Eq.~\ref{Dgen} will be
recognized as Fourier transforms, allowing us to benefit from the
efficiency of the FFT algorithm.  The element-by-element
ratio of these two, $s_p = N_p / D_p,$ forms the filtered signal
map.

\subsection{Error propagation}

The error in our estimate of $s_p,$ which we denote $n_p$, is the rms
deviation of the best fit PSF amplitude, $s_p$, from the actual
amplitude, $a_p$, of a point source that is centered on the $p$'th
pixel. i.e.\ $n_p^2 = \langle | s_p - a_p |^2 \rangle.$ To estimate
$n_p^2,$ we consider the influence of a point source
located at the $p$'th map pixel on a neighboring pixel $i.$ 
\begin{equation}
d({\bf x_i}) = a_p f({\bf x_i} - {\bf x_p}) + \sqrt{C_{ii}}.
\label{pix_val}
\end{equation}
As mentioned above, we recognize that the possibility of other
resolved sources being quite close to pixel $i$ (source blending) is
ignored in Eq.~\ref{pix_val} and must be treated iteratively.  Using
Eq.~\ref{pix_val} for $d({\bf x_i})$ and Eq.~\ref{s_p} for $s_p,$ our
estimate for $n_p^2$ is
\begin{equation}
n_p^2 = \langle | s_p - a_p |^2 \rangle = 
\frac{\displaystyle{
\sum_{k,l} \sum_{t,u}
W_{kl} W_{tu} C_{kt}
f({\bf x_l} - {\bf x_p}) f({\bf x_u} - {\bf x_p})
}}
{\displaystyle{ \left(
\sum_{k,l} W_{kl}
f({\bf x_k} - {\bf x_p}) f({\bf x_l} - {\bf x_p})
\right)^2 } },
\label{err}
\end{equation}
and since $W =C^{-1},$ the above expression simplifies to
\begin{equation}
n_p = {1 \over 
\displaystyle{\sqrt{ \sum_{k,l} W_{kl}
f({\bf x_k} - {\bf x_p}) f({\bf x_l} - {\bf x_p})
} } } = {1 \over \sqrt{D_p}}.
\label{n_p}
\end{equation}
Thus, when the optimal $W$ is used, no extra steps are needed to
evaluate $n_p,$ as $D_p$ already exists from generating the filtered
signal map.  

\subsection{Source significance and goodness of fit}

The filtered signal-to-noise map, filled with values $s_p/n_p,$ can be
used to both identify {\em and locate} sources of high significance.
Rewriting Eq.~\ref{chisq1} in a more convenient form using
Eq.~\ref{s_p} and Eq.~\ref{n_p} gives
\begin{equation}
\chi_p^2 = \sum_{k,l} W_{kl} d({\bf x_k}) d({\bf x_l}) - (s_p/n_p)^2.
\label{chisq2}
\end{equation}
Because the first term in Eq.~\ref{chisq2} is common to all pixels
$p,$ the second term is a direct indication of the goodness of fit at
the $p$'th pixel.  Thus, the locations of peaks in the signal-to-noise
map, $s_p/n_p$, provide the best estimates of point-source positions
on the map.

\section{The case of uniform coverage}
\label{simple_filter}

It is instructive to first consider the case of uniform coverage.  In
this case, $\epsilon({\bf x_k})=\epsilon({\bf x_l})={\rm cnst}$ in
Eqs.~\ref{Ngen} and~\ref{Dgen} and so can be moved outside the
summations.  The sums over $k$ and $l$ are then recognized as Fourier
transforms (FTs) of the signal map ($d({\bf x})$) and the PSF ($f({\bf
  x})$), leading to the result that
\begin{equation}
s_p = \frac{
\displaystyle{\sum_{a} {\tilde{f}^*({\bf k_a}) \tilde{d}({\bf k_a}) \over
    V^2({\bf k_a)}}
~\mathrm{exp} [ 2 \pi j {\bf k_a} \cdot {\bf x_p} ]}
}
{\displaystyle{
\sum_{a} { |\tilde{f}({\bf k_a})|^2 \over  V^2({\bf k_a}) }}
},
\label{sp_simple}
\end{equation}
where $\tilde{f}({\bf k_a})$ and $\tilde{d}({\bf k_a})$ are the 2-d
FTs of $f({\bf x_i})$ and $d({\bf x_i})$ respectively.  The above
expression is the familiar ``matched filter'' that is commonly used in
mm/sub-mm astronomy for identifying point sources \citep[for example,
  see][]{Tegmark1998, Barreiro2003, Vio2004, Barnard2004, Chapin2011}
and is essentially a band-pass filter in the sense that the data
($\tilde{d}$) are low-pass filtered by the PSF $\tilde{f}$
and high-pass filtered by $1/V^2$ as the PSD usually
displays a $1/f$-type trend with spatial frequency.  The denominator
of Eq.~\ref{sp_simple} simply provides the correct normalization.
What Eq.~\ref{sp_simple} shows is that the point-source finding
technique of fitting a PSF to each map pixel converges to the
conventional matched filter in the case of uniform coverage.

Another important feature of Eq.~\ref{sp_simple} is that, in terms of
the FTs $\tilde{f}({\bf k_a})$ and $\tilde{d}({\bf k_a})$ the
numerator and the denominator are each reduced to a single sum of
$N_\mathrm{pixel}$ terms and the total calculation scales as
$N_\mathrm{pixel}\log_2N_\mathrm{pixel}$ rather than as
$N_\mathrm{pixel}^2$.  As for having to evaluate $s_p$ at each pixel,
we note that (1) the denominator needs to be evaluated just once for
all pixels and (2) a ``map'' of the numerator is simply an inverse
Fourier transform (IFT).  Therefore, the entire filtering process can
be accomplished with $\sim {\cal
  O}(N_\mathrm{pixel}\log_2N_\mathrm{pixel})$ calculations.  On a
desktop computer with a 2.5\,GHz Intel processor and 4.7 GB of RAM,
the $\sim 200,000$ pixel map of Fig.~\ref{raw_maps} can be filtered
according to Eq.~\ref{sp_simple}, using the IDL programming language,
in under 0.5 seconds.

Using the standard matched filter is similar to performing a
simple least squares minimization under the approximation that all pixels in
the map have the same uncertainty.  If the underlying noise is {\em not} uniform
and this method is used, one must recognize that $W \ne C^{-1}$ and so the
error distributions of the $s_p$ cannot be formally calculated as given
in Eq.~\ref{n_p} above.

\section{The case of non-uniform coverage}
\label{general_filter}

We show here that, in a map with non-uniform coverage, it is possible
to treat the general problem of point source identification in a
computationally efficient, yet mathematically sound way.  We find that
loosening the assumption of uniform coverage results in an increase in
computation time that is noticeable but inconsequential in practical
terms.  As the calculation of $D$ turns out to be more complex than
the calculation of $N,$ we tackle $N$ first.

\subsection{The calculation of $N$}

$N,$ as expressed in Eq.~\ref{Ngen}, can be calculated through a
sequence of FTs and IFTs.  We motivate this by casting Eq.~\ref{Ngen}
in a more suggestive manner:
\begin{equation}
N_p = 
{1 \over N_\mathrm{pixel}}
\sum_k {f({\bf x_k} - {\bf x_p}) \over \epsilon({\bf x_k})}
\left( \sum_a \frac
{ e^{ 2 \pi j {\bf k_a} \cdot {\bf x_k} } }
{V^2({\bf k_a})}
\left( {1 \over N_\mathrm{pixel}} \sum_{l}
{d({\bf x_l}) \over \epsilon({\bf x_l})} e^{-2 \pi j {\bf k_a} \cdot {\bf x_l}}
\right)
\right)
\label{Nfinal}
\end{equation}
Proceeding from the innermost parentheses outward, the steps needed to
calculate $N$ are:
\begin{enumerate}
\item
Form a new map $M,$ filled with values $M({\bf x_l}) = d({\bf x_l}) /
\epsilon({\bf x_l}).$
\item
The calculation within the innermost parentheses can be recognized as
a Fourier Transform (see Appendix A).  Therefore, form $\tilde{M} =
\mathrm{FT}[M].$
\item
Similarly, the calculation within the next set of parentheses is an
IFT.  Thus, form $P = \mathrm{IFT}[\tilde{M}/V^2].$
\item
Form a new matrix $Q,$ which is filled with values $Q({\bf x_k}) =
P({\bf x_k}) / \epsilon({\bf x_k}).$ The sum over $k$ in
Eq.~\ref{Nfinal} is simply a convolution of $Q$ by the PSF $f.$
\item
As it is computationally advantageous to carry out this convolution in
the Fourier domain, calculate $\tilde{f} = \mathrm{FT}[f]$ and
$\tilde{Q} = \mathrm{FT}[Q].$
\item
Finally, $N = \mathrm{IFT}[\tilde{f}^* \tilde{Q}].$
\end{enumerate}
There are 5 FFTs in this sequence.  On the same standard desktop PC
described in Section~\ref{simple_filter}, the complete computation sketched
above takes less than one second for a $\sim 200,000$-pixel map.

\subsection{The calculation of $D$}

The calculation of the denominator, $D$, is more involved than
the numerator as there are no simplifications that allow $D$ to reduce
to a series of Fourier Transforms.  As a result, a full calculation of
$D$ requires of ${\cal O}(N_\mathrm{pixel}^2)$ calculations.  In the
case of the $\sim 200,000$-pixel map used here, this takes about seven
hours on the basic computing platform considered in
Section~\ref{simple_filter}.  We can do better though.  Below we give
an approximation to $D$ with an accuracy better than 0.1\% that takes
$\sim 2$ minutes to calculate.

Our Approximation of $D$ makes use of the band-diagonal nature of the
noise correlation matrix or, equivalently, the smoothly varying nature
of $V^2({\bf k_a})$ as discussed in section \ref{outline}.  We start
by defining a new vector, ${\bf x_d} = {\bf x_k} - {\bf x_l},$ and
re-writing Eq.~\ref{Dgen} as
\begin{equation}
D_p =
{1 \over N_\mathrm{pixel}^2}
\sum_d Z({\bf x_d})
\sum_l
\frac
{f({\bf x_l} - {\bf x_p} + {\bf x_d}) f({\bf x_l} - {\bf x_p})}
{\epsilon({\bf x_l} - {\bf x_d}) \epsilon({\bf x_l})},
\label{D2}
\end{equation}
where $Z({\bf x_d})$ is the IFT of $1/V^2({\bf k_a}),$ or
\begin{equation}
Z({\bf x_d}) = 
\sum_a \frac
{ e^{ 2 \pi j {\bf k_a} \cdot {\bf x_d} } }
{V^2({\bf k_a})}.
\label{Zxd}
\end{equation}
To clearly demonstrate that the sum over $l$ is a convolution,
we define the new functions $F_d$ and $R_d$ as follows:
\begin{eqnarray}
F_d({\bf x}) & = & f({\bf x} + {\bf x_d}) f({\bf x}) \\
R_d({\bf x}) & = & {1 \over \epsilon({\bf x} + {\bf x_d}) \epsilon({\bf x})}.
\end{eqnarray}
Thus, $F_d$ (or $R_d$) is the product of $f$ (or $1/\epsilon$) and a
version of $f$ (or $1/\epsilon$) that is shifted by the vector ${\bf
  x_d}.$ Using these definitions, we may express $D_p$ as
\begin{equation}
D_p =
{1 \over N_\mathrm{pixel}^2}
\sum_d Z({\bf x_d})
\sum_l
F_d({\bf x_l} - {\bf x_p}) R_d({\bf x_l}),
\end{equation}
which is recognizable as a convolution.  Finally, using the discrete
convolution theorem (see Eq.~\ref{app_conv}), we find that
\begin{equation}
D_p = {1 \over N_\mathrm{pixel}}
\sum_d Z({\bf x_d})
\left(
\sum_a
\tilde{F_d}^*({\bf k_a}) \tilde{R_d}({\bf k_a}) e^{-2 \pi j {\bf k_a}
  \cdot {\bf x_p}}
\right).
\label{Dfinal}
\end{equation}

The advantage of casting $D$ in the form of Eq.~\ref{Dfinal} is the
following: $Z({\bf x_d})$ is narrowly peaked near ${\bf x_d = 0}$
because it is the IFT of a smoothly varying function $1/V^2({\bf
  k_a}).$  Therefore, the sum over $d$ converges
very rapidly at small ${\bf x_d}$ and changes very little at large
${\bf x_d}$ and so high accuracy in the calculation of $D$ can still
be obtained after truncating the calculation at $N_t\ll N_{\rm pixel}$
terms.  The choice of $N_t$ will depend on the properties of $Z({\bf
  x_d})$ and the desired accuracy in $D$.  For example, for our AzTEC
maps we find that if we limit the sum over $d$ to the $N_t =
N_\mathrm{pixel}/200 \approx 1,000$ terms where $| Z({\bf x_d})|$ is
largest, $D$ converges to within 0.1\% of its final form in useful
parts of the map (the $>5$\% coverage region) and requires only 2
minutes to compute.

The steps we follow to generate our approximate $D$ are then
\begin{enumerate}
\item
Compute $Z = \mathrm{IFT}[1/V^2]$ and find the $N_t$
positions ${\bf x_d}$ where $|Z|$ is largest.  Then, for each of those
${\bf x_d},$
\item
shift $f$ by ${\bf x_d}$ and multiply by the unshifted $f$ to
form $F_d.$ Then generate $R_d$ by performing the same steps on
$1/\epsilon;$
\item
compute the Fourier Transforms $\tilde{F_d} = \mathrm{FT}[F_d]$ and $\tilde{R_d} =
\mathrm{FT}[R_d];$
\item
compute the term within parentheses in Eq.~\ref{Dfinal}
as $G({\bf x_d}) = \mathrm{IFT}(\tilde{F_d}^* \tilde{R_d});$
\item
Repeat for the $N_t$ vectors (${\bf x_d}$) chosen, and sum the terms
$Z({\bf x_d}) G({\bf x_d})/N_\mathrm{pixel}.$
\end{enumerate}

\section{Application of the optimal filter to AzTEC maps}
\label{application}

Here, we will demonstrate how the methods developed above have been
applied to AzTEC maps, using the field of Fig.~\ref{raw_maps} as an
example.  Although the examples and justifications presented here are
based on AzTEC data, we note that many of the trends and techniques
identified here are typical of most observations and analysis chains.

\subsection{Validation of assumptions using the PSF}

In AzTEC, we have always used an accurate simulation of the PSF,
rather than a generic form such as a Gaussian, as the template $f({\bf
  x})$ used in the filter.  According to the methodology developed in
section \ref{outline}, this is the correct template to use, and should
lead to higher accuracy of filtered maps.  The simulation used for
generating the PSF includes effects such as individual detector beam
shapes, the location and orientation of the field during each
observation, as well as ``cleaning'' and filtering steps identical to
those used on the true field.  The details of how the PSF is generated
in the AzTEC data analysis can be found in \citet{Downes2012} and
\citet{Scott2008}.
\begin{figure}[h!]
\centering
\leavevmode
\begin{tabular}{cc}
\includegraphics[height=2.5in]{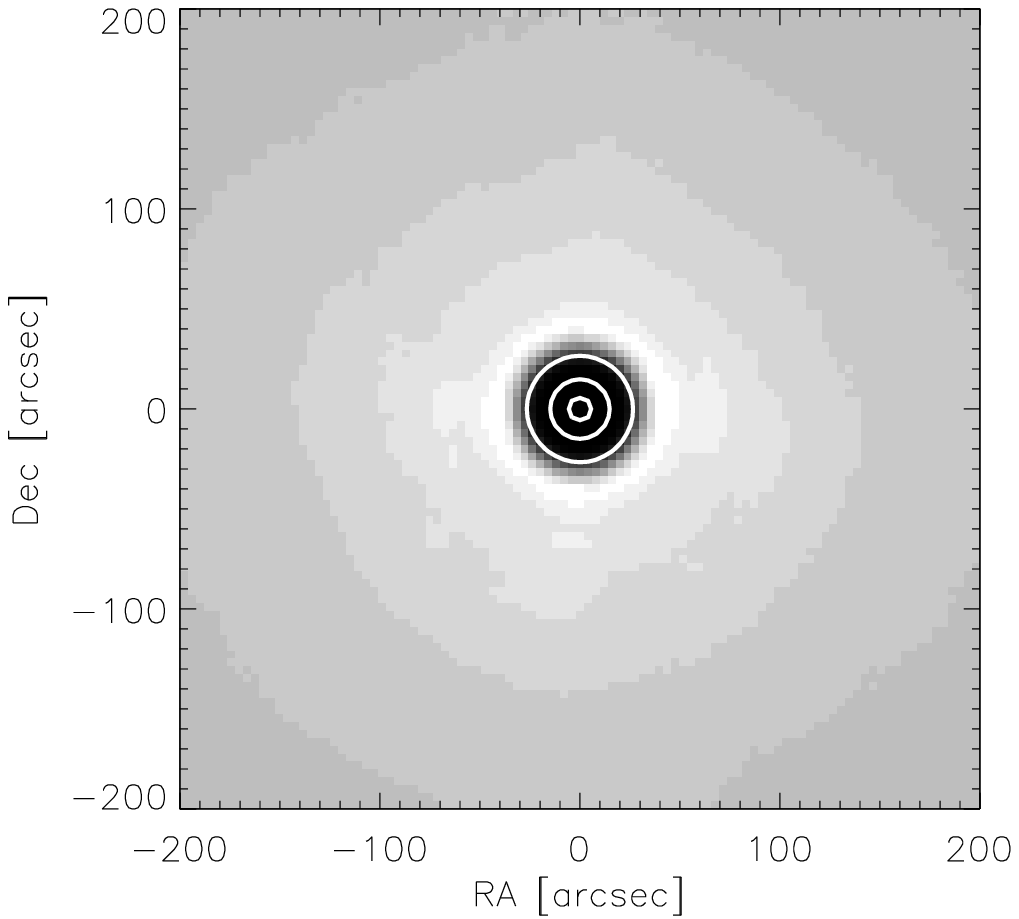}
& 
\includegraphics[height=2.5in]{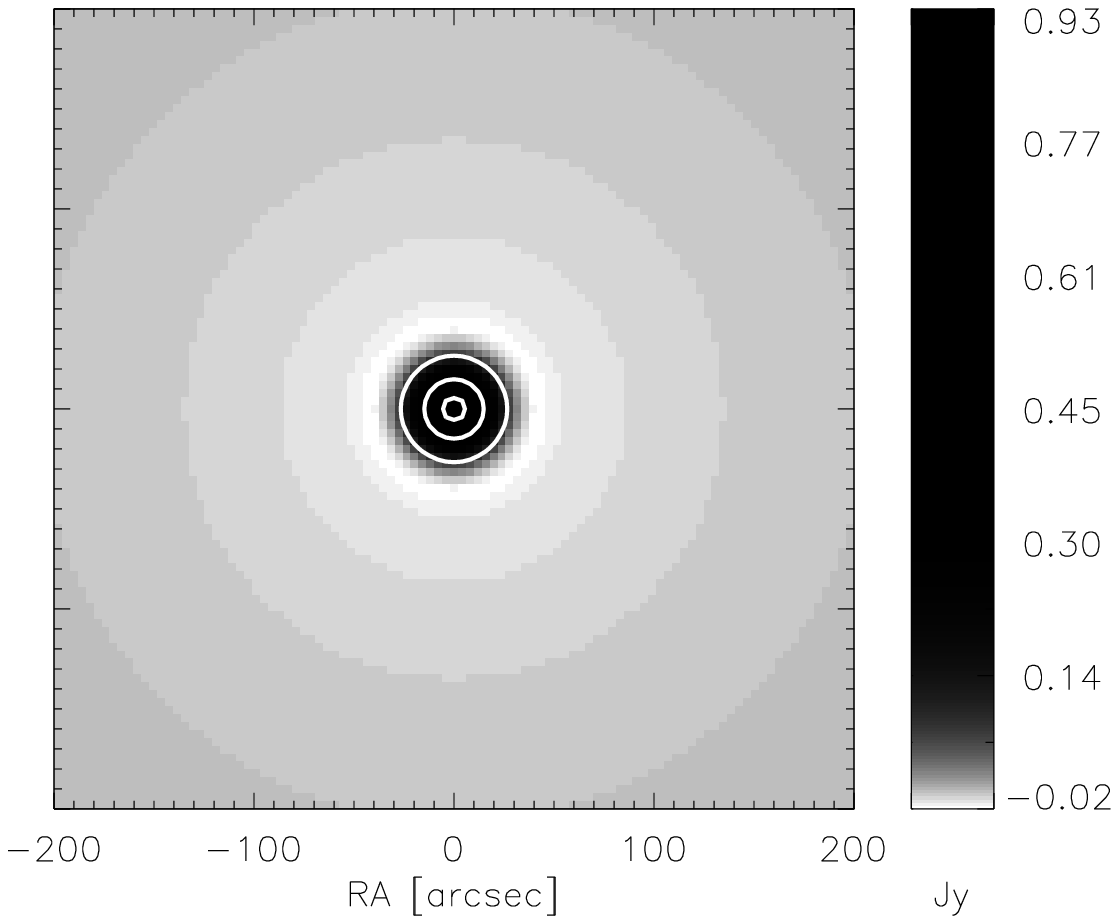}\\
(a) & (b)
\end{tabular}
\caption{(a) The PSF resulting from optics and data ``cleaning.''  The
  grey scale is stretched to emphasize the asymmetric and, at times
  negative, low-signal values.  The dark high-signal regions follow a
  rotationally symmetric profile.  The three white contours are at
  10\%, 50\%, and 90\% of the peak value.  The peak value is 0.93\% of
  the inserted source brightness, indicating the amount by which a
  point source is attenuated due to the ``cleaning'' of atmospheric
  contamination.  (b) The rotationally symmetrized version of (a) used
  as the fit template.  The central region has a FWHM of $\sim
  29^{\prime\prime}.$
\label{kernel}
}
\end{figure}
Fig.~\ref{kernel}(a) shows the PSF that applies to the AzTEC map of
Fig.~\ref{raw_maps}.  Strictly speaking, the PSF of
Fig.~\ref{kernel}(a) only applies to a point source at a particular
location on the map.  Therefore, before adopting assumption 1 of
section \ref{outline:assumptions}, we constructed several PSFs that apply to
different locations on a map and verified that they are all very
similar (at the few percent level) in terms of amplitude, width, and
general shape \citep{Scott2008}.  The only noticeable difference
between these PSFs is the orientation of the low-level asymmetric
features seen in Fig~\ref{kernel}(a).  Therefore, we use a
rotationally symmetrized version of the PSF, shown in
Fig.~\ref{kernel}(b), as the filter template.  To evaluate the
inaccuracy introduced by rotationally symmetrizing the PSF, we have
fit the PSF of Fig.~\ref{kernel}(b) to the PSF of
Fig.~\ref{kernel}(a), and find that the best fit amplitude differs
from the actual amplitude of Fig.~\ref{kernel}(a) by $\sim0.5$\%,
which is small compared to other errors we expect.  Therefore, we
adopt assumption 1 and the rotationally symmetrized PSF of
Fig.~\ref{kernel}(b).

Assumption 2 of section \ref{outline:assumptions} would be difficult
to adopt if the field is imaged only once, so that the field has an
essentially fixed Az-El orientation during the observation and each
point on the map is scanned over in essentially one direction.
However, AzTEC and other mm/sub-mm surveys routinely adopt assumption
2 because each of the mapped fields is imaged many times with varying
orientations and scan directions.  This is evident in
Fig.~\ref{kernel}(a) where the grey scale has been stretched to
highlight the low-level asymmetric features of the PSF.  We believe
that such features are due to a small anisotropy in the final
aggregate of scan directions from the 45 observation of this field.
However, the fact that these features are so faint compared to the
symmetric central parts (refer to the color bar of Fig.~\ref{kernel})
supports our adoption of assumption 2.

\subsection{Estimation of the PSD}

In the AzTEC data analysis, the PSD corresponding to an observation is
evaluated using Fourier transforms (FTs) of noise realization maps.
Our noise realizations are generated using a ``jackknife'' method,
where the signs of the detector time-streams are switched many times
for each observation prior to map making \citep{Scott2008}.  An
important consideration is to use long-enough time lags between the
random sign switchings in order to preserve long-range noise features
(low-frequency map noise) due to residual atmospheric contamination
and instrumental drifts.  The $\sim10\,$s time scales used are longer
than the scan turn-around times but short enough that astronomical
signal will not show up in noise realizations.  Next, the noise
realization time-streams are put through the same cleaning and
filtering steps as the actual data, before maps are made.  For a map
such as Fig.~\ref{raw_maps}, we generate 100 independent noise
realization maps.

Because noise realizations are used for generating the PSD, it is free
from astronomical signal and only includes the two dominant noise
sources, residual atmospheric contamination and instrumental noise.
The sub-dominant yet noticeable contribution of confusion noise is
left out from the PSD for convenience.\footnote{We note that other
  analyses that make use of the filtered signal map, such as
  estimating de-boosted source fluxes, stacking analyses, and
  number-counts estimates will not be skewed by the omission of
  confusion noise due to the use of source realizations and/or the
  AzTEC PSF having zero mean \citep[see][]{Scott2008}.  On the other
  hand, source lists for follow-up observations, which are based
  purely on $s/n$ will be very slightly biased due to the omission of
  confusion noise.  This bias would favor peaks found in regions that
  have an over-density of unresolved sources, which can be beneficial
  to our understanding of SMGs and their environments.}  However, the
method presented here, of using a PSD to characterize noise
correlations, can fully accommodate astronomical ``noise sources''
such as confusion noise.  For instance, \citet{Chapin2011} describe
how confusion noise was included in their matched filter.  In general,
astronomical effects such as confusion noise are frequency dependent.
At longer wavelengths, a method for including the CMB within a matched
filter is described in \citet{Tegmark1998}.

\begin{figure}[h!]
\centering
\includegraphics[width=5in]{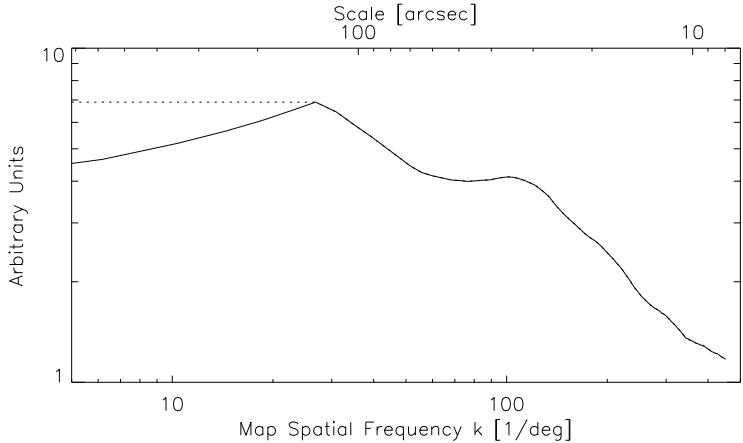}
\caption{The average power spectral density (PSD) of noise in
  noise-realization maps.  The broad peak at the PSF scale ($\sim
  30^{\prime\prime}$) is believed to be caused by atmospheric emission
  that is imaged in between the random sign switches used to generate
  noise realizations.  The dotted line indicates an alteration to the
  PSD (of plateauing at low frequency) that we have tried.
\label{psd}
}
\end{figure}
In our case, the noise PSD is generated by truncating each noise map
to include only the central $>70$\% coverage region (see
Fig.~\ref{raw_maps}(b)).  Then, the 2-d FTs obtained from these
regions are rotationally symmetrized and averaged to obtain the PSD
shown in Fig.~\ref{psd}.  As expected, the noise increases with
decreasing $k$ over most of the $k$ range.  A broad peak is visible at
$\sim 30^{\prime\prime},$ which is close to the FWHM of the PSF.
This, we believe, is caused by the optical imaging of atmospheric
fluctuations in between the random sign switches used to generate
noise realizations.  Near the $\sim 100^{\prime\prime}$ scale, the PSD
turns over and starts to decrease.  This decline cannot be a map-size
effect because the $>70$\% coverage region used to construct the PSD
extends $>700^{\prime\prime}$ in all directions.  We believe that the
PSD's turn-over is real and that it is caused by the principle
component analysis (PCA) based cleaning that the data is put through
prior to map making, in order to mitigate the effects of atmospheric
contamination and detector cross-talk.  As PCA cleaning makes use of
detector-detector time-stream correlations to ``subtract'' out these
effects, it makes sense that the PSD shows a decline on the scale of
the detector array (the spacing of individual detector beams is $\sim
40^{\prime\prime}$ and the footprint of the entire array is $\sim
480^{\prime\prime}$ for AzTEC-ASTE).  To test if this long-range
decline in the PSD is indeed real, we have performed the optimal
filter presented here using the measured PSD as well as a modified PSD
that plateaus at small $k,$ as indicated by the dotted line of
Fig.~\ref{psd}.  The fact that there is no perceptible difference in
results indicates that the map noise power is indeed low at low $k.$
As noted earlier, an important property of the PSD of Fig.~\ref{psd}
is that it is a smoothly varying function, as opposed to a narrowly
peaked one.

\subsection{Filtering of maps}
\label{filt_of_maps}

\begin{figure}[h!]
\centering \leavevmode
\begin{tabular}{cc}
\includegraphics[height=2.9in]{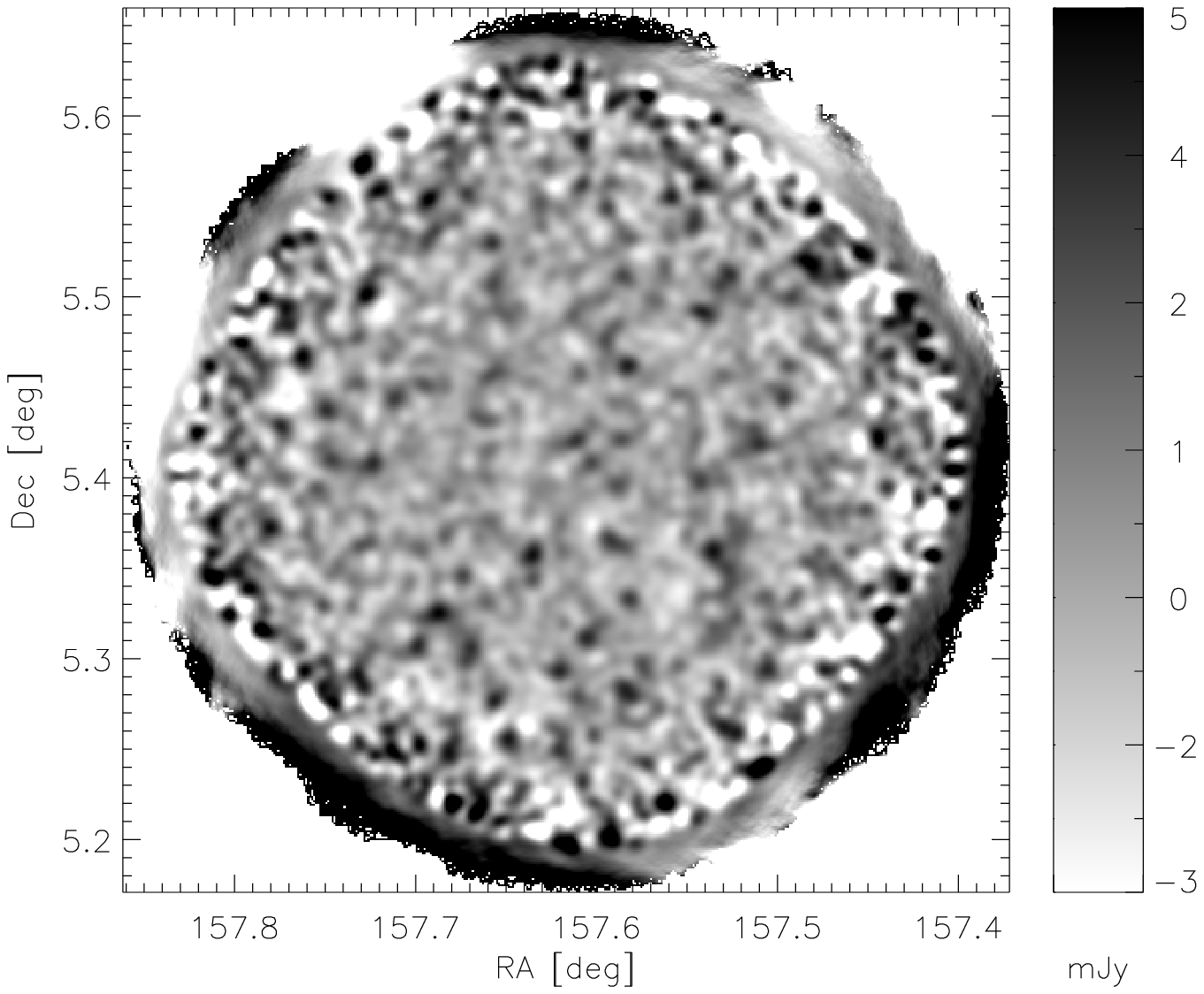}
& 
\includegraphics[height=2.9in]{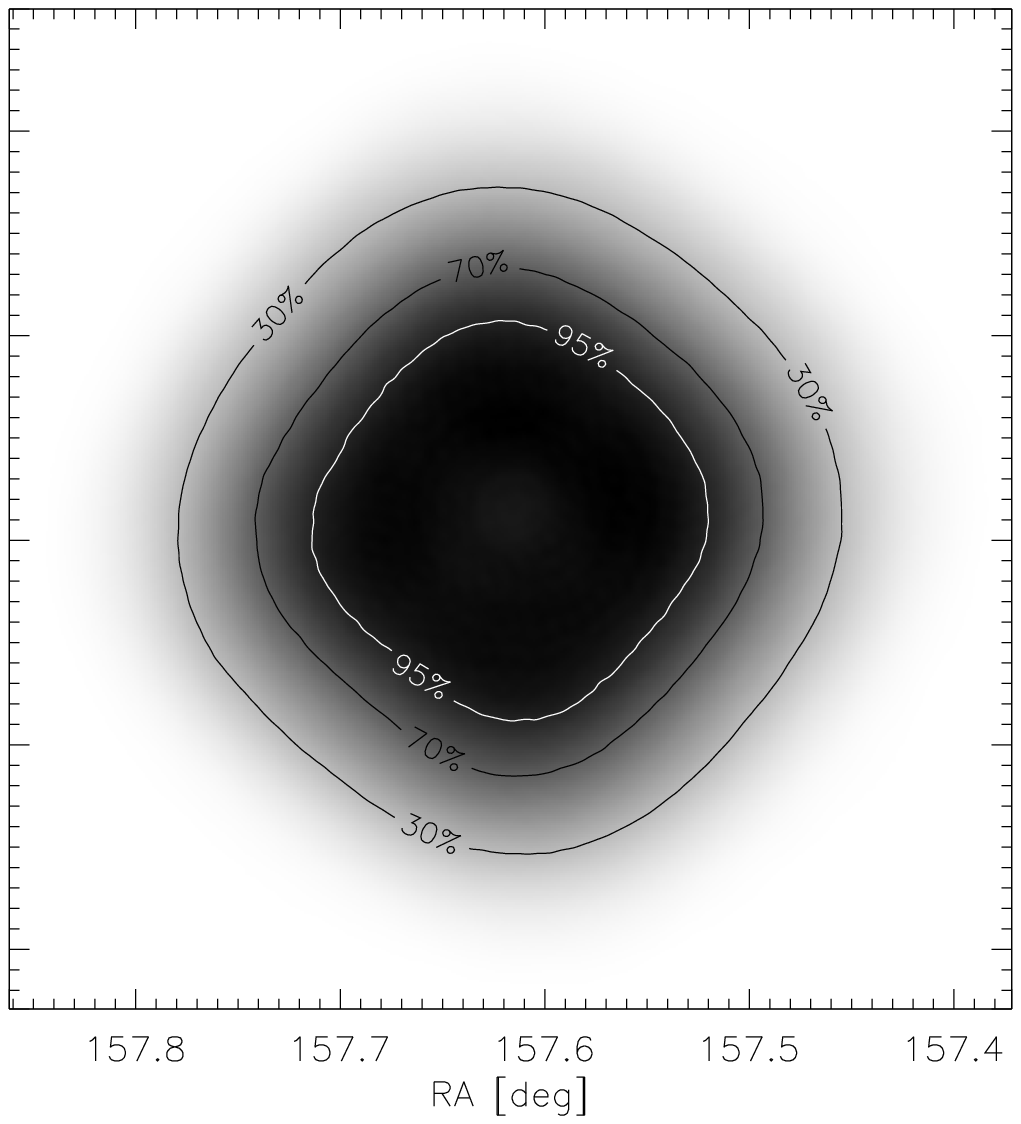}\\
(a) & (b)
\end{tabular}
\caption{Filtered signal map and filtered coverage map of the entire
  field, generated according to the optimal filter of section
  \ref{general_filter}.
\label{fig_general_filter}
}
\end{figure}

Once the PSF and PSD are available, the methods of section
\ref{general_filter} can be applied to obtain the filtered signal and
coverage maps.  In practice, more time is spent on the calculation of
$N$ than the calculation of $D$ during the filtering process, even
though $D$ involves a more complex calculation (see section
\ref{general_filter}).  This is because $N$ needs to be evaluated
separately for the actual map and each noise realization, as $N$
depends on $d({\bf x})$ (see Eq.~\ref{Ngen}), while $D$ needs to be
evaluated only once for all 101 maps.  For the example field used, the
filtered signal and coverage maps are shown in
Fig.~\ref{fig_general_filter}.

\begin{figure}[h!]
\plotone{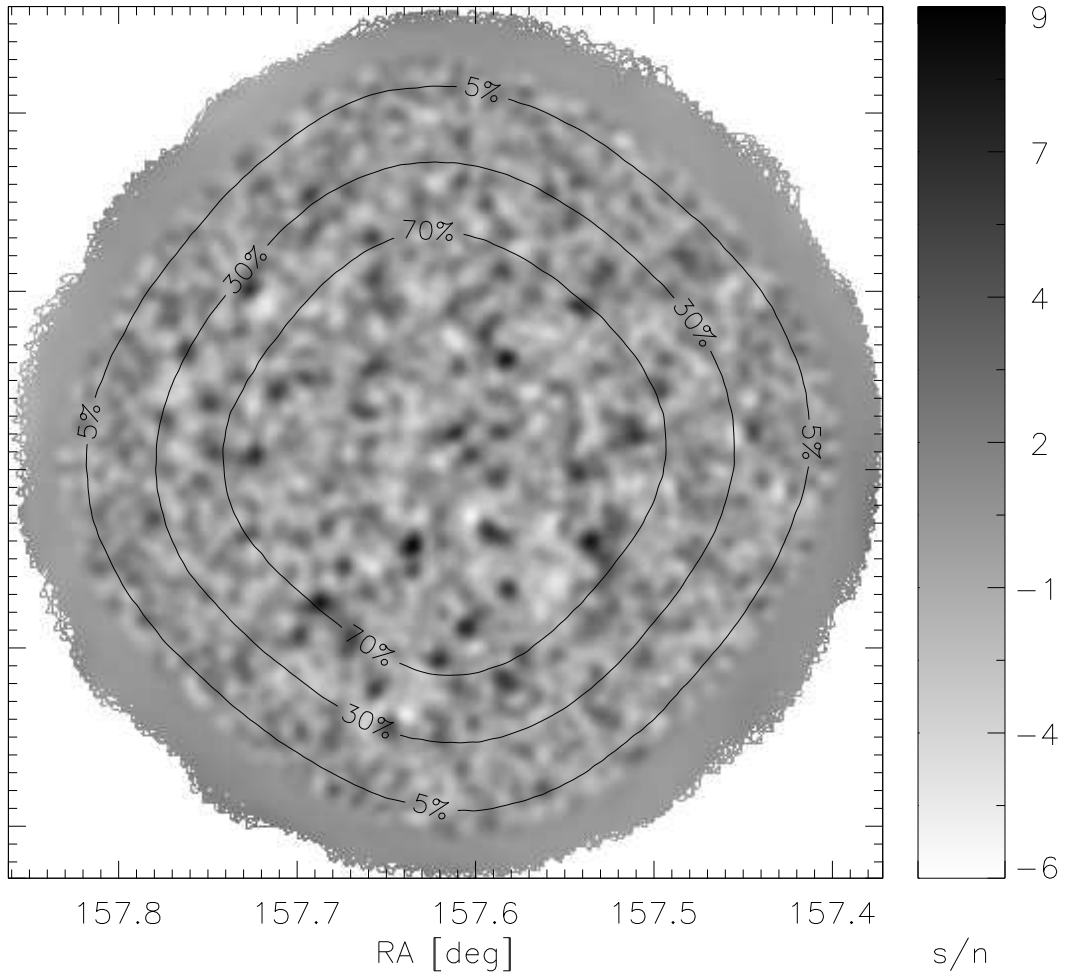}
\caption{The filtered signal-to-noise map.  The 5\%, 30\%, and 70\%
  coverage contours are overlayed.
\label{sn}
}
\end{figure}

The filtered coverage map must be filled with values $1/n_p^2,$ and
therefore, according to Eq.~\ref{n_p}, it is simply equal to the map
$D.$ But $D$ is generated by propagating the initial noise variances,
$\epsilon({\bf x_p}),$ according to Eq.~\ref{Dgen}.  Thus the noise
estimates contained in $D$ are only as good as the starting noise
estimates $\epsilon({\bf x_p}),$ which are computed from estimates of
individual detectors' noise levels ``near the time'' that they
contribute a sample to the map pixel in question \citep[see][for
  details]{Scott2008}.  Therefore, before proceeding to the step of
identifying point sources, we check the accuracy of noise estimates
contained in $D$ using noise realization maps.  First, we form a
second noise estimate $n^\prime_p$ for each pixel by taking the
standard deviation of the 100 noise-map values found at that pixel.
Next, we form a second coverage map $T$ filled with the values
$1/{n^\prime_p}^2.$ Although $T$ provides a robust estimate of
coverage, it is noisier than the the original coverage map $D$ due to
the sample size of 100 (we expect a $\sim$7\% error in $T_p$ for this
sample size).  Despite the noisy appearance of $T$, its overall shape
agrees very well with $D.$ However, $T$ and $D$ often differ from each
other by an overall scaling factor.  We determine this scaling factor
by comparing the average values of $T$ and $D$ in the $>$70\% coverage
region and applying a correction factor to $D$ in order to obtain the
final filtered coverage map.  Thus, the filtered coverage map is $\eta
D,$ where
\begin{equation}
\eta = {\langle T \rangle_{70\%} \over \langle D \rangle_{70\%}},
\end{equation}
and $\eta$ usually lies in the range 0.85-0.95, depending on the
field.  Although we use just the $>$70\% coverage region to find
$\eta,$ the agreement between $T$ and $\eta D$ remains good out to the
very edge of the mapped region until the coverage dips below 5\%.
Beyond this region, $\eta D$ is consistently larger than $T.$
Therefore, we do not trust the filtered coverage map beyond the 5\%
region and exclude this area when searching for point sources.
Finally, in Fig.~\ref{sn}, we present the filtered signal-to-noise
map, which is used directly to find point sources.  It is generated by
dividing the filtered signal map ($N_p/D_p$) by the filtered noise map
($1/\sqrt{\eta D_p}$).\footnote{The correction factor $\eta$ does not
  need to be applied in generating the filtered signal map because $N$
  and $D$ both contain two factors of $\epsilon$ (see Eq.~\ref{Ngen}
  and Eq.~\ref{Dgen}).}

\section{Results and discussion}
\label{conclusion}

It is encouraging that the signal-to-noise map of Fig.~\ref{sn} has
its most prominent peaks near the center, where the coverage is
highest, and that the number of peaks as well as their amplitude
declines smoothly toward the outer regions.  Thus, even though the
filtered signal map has large fluctuations near the edges (see
Fig.~\ref{fig_general_filter}(a)), the noise in those regions are
accounted for in this optimal filter and there in no need for a
by-hand coverage cut to eliminate spurious behavior near the edges.
Throughout the AzTEC data analysis campaign, which has led to the
publication of many source lists, variations of the filter presented
here have been used.  Therefore, the effectiveness of this technique
may be assessed by the success rate of follow-up observations.  In
this regard, AzTEC has enjoyed an excellent record thus far.  For
instance, of 15 AzTEC point sources followed up by the Submillimeter
Array (SMA), all were successfully observed
\citep{Younger2007,Younger2009}. This is a very good record in
comparison to the usual success rate of mm/sub-mm follow up studies.

In terms of weighting a fit according to all possible sources of
uncertainty, the filtering technique presented in section
\ref{general_filter} (filter A) is clearly closer to optimal than the
standard matched filter discussed in section \ref{simple_filter}
(filter B), especially in the presence of significant coverage
gradients.  However, filter A contains a larger number of steps and
computations.  In terms of implementation, we have shown that the
intricacies of filter A have virtually no effect on the total
computing time devoted to a field.  However, converting these
additional steps into functional computer code does require more time
and effort.  Therefore, it is fair to ask {\em how much} better filter
A performs compared to filter B in a practical application.  Of
course, the answer depends on the specifics of the application.  For
instance, the acuteness of coverage gradients in a map will be
important for this assessment.  We will try to answer this question in
the context of the typical AzTEC map that we have used here for
illustrative purposes.

First, when a fit is closer to optimal, it yields smaller error bars
on the fit parameters than a less optimized fit.  A fair way of
comparing error bars here is to compare the $T$ maps (generated using
noise realizations, as described in section \ref{filt_of_maps}) that
result from the two filters.  We find that the post-filter coverage
($1/n_p^2$) is on average 5-6\% better with filter A than with
filter B over most of the map, including the $>70$\% coverage region.
Thus, filter B is more prone to error, which in this context leads to
higher rates of false detections as well as non-detections of truly
significant sources.

\begin{deluxetable}{lccc}
\tablecaption{The number of peaks with $s/n > 3.5$ found with the
  fully optimal filter of section \ref{general_filter} (filter A) and
  the standard matched filter of section \ref{simple_filter} (filter
  B) are reported.  The last column gives the number of ``sources''
  found with filter B but not with filter A.  The 30-70\% coverage
  region has an area that is 75\% of the $>$70\% coverage
  region.\label{tb1}}
\tablewidth{0pt}
\tablehead{
\colhead{Map region} & \colhead{Filter A} & \colhead{Filter B} & 
\colhead{B but not A}
}
\startdata
$>$70\% coverage & 31    &   31  &  2\\
30-70\% coverage & 15    &   18  &  3\\
\enddata
\end{deluxetable}

To form an impression regarding these rates, we summarize in
table~\ref{tb1} the results obtained with each filter in different
regions of the map.  The ``source candidates'' reported in
table~\ref{tb1} are defined to be peaks in the signal-to-noise map
(Fig.~\ref{sn}) that exceed a value of 3.5.  This is a good choice of
threshold because, on average, pure noise realizations yield only
$\sim 1$ noise peak with signal/noise $> 3.5$ over the entire map.
Given the non-uniformity of coverage within the 30-70\% region, it is
not surprising that filter B ``detects'' 3 peaks that are judged by
filter A (more accurately) to have $s/n<3.5.$ It is interesting that
even in the $>70$\% coverage region, the source lists generated with
filters A and B differ by two sources.  The results reported in
table~\ref{tb1} provide some grounds for the reader to gauge whether
the additional work involved in implementing filter A is worth the
effort.  Of course, the optimal filter presented here will be most
useful when the coverage of a field has large non-uniformities, unlike
in the example field used here.

Thus far, many AzTEC publications only provide source lists from map
regions with $>70$\% coverage.  As the optimal filter presented here
is geared to handle coverage non-uniformities, we believe that the
reliable region for source extraction can be extended to lower
coverage thresholds.  As mentioned in section \ref{filt_of_maps},
regions with $< 5$\% coverage should be excluded.  In fact, it may be
best to exclude a larger region.  For instance, not all elements of
the detector array have imaged the outer regions of the map.
Therefore, those regions may be systematically biased in some way
compared to regions viewed and appraised by all detectors.  Given that
the AzTEC-ASTE detector array has a foot print on the order of
$480^{\prime\prime},$ it is reasonable to leave out a border of half
that size, and use the $> 30$\% coverage region of Fig.~\ref{sn} for
source identification.  This represents an increase of 75\% in the
source extraction area.  Table~\ref{tb1} shows that $\sim 50$\% more
source candidates can be obtained this way.

The optimal filter described here is now part of the AzTEC data
analysis pipeline, which may be downloaded (by anonymous ftp) \\ at
\url{http://www.astro.umass.edu/aztec/Software/software.html}.  Within
this suite, the particular IDL routine that implements the optimal
filter \\ is {\em aztec\_adapative\_wiener\_filter.pro.}

\acknowledgments

This work has been funded, in part, by NSF grant AST-0907952.  KSS is
supported by the National Radio Astronomy Observatory, which is a
facility of the National Science Foundation operated under cooperative
agreement by Associated Universities, Inc.

\appendix

\section{Appendix: Fourier transform conventions}

In this work, $\mathrm{FT}[g]$ denotes the discrete 2-dimensional Fourier
transform of the function $g({\bf x}).$  Following the convention used
by IDL and other high-level programing languages, this Fourier
transform is defined as
\begin{equation}
\tilde{g}({\bf k}) = \tilde{g}(k_a \hat{\imath} + k_b \hat{\jmath}) =
{1 \over N_x N_y}
\sum_{m = 0}^{N_x -1}
\sum_{n = 0}^{N_y -1}
g(x_m \hat{\imath} + y_n \hat{\jmath})
e^{- 2 \pi j a m / N_x }
e^{- 2 \pi j b n / N_y },
\end{equation}
where $N_x$ and $N_y$ are the number of pixels in the $x$ and $y$
dimensions.  Therefore, $N_x N_y = N_\mathrm{pixel}.$ The $x_m$ and
$y_n$ above, take on the $x$ and $y$ values of all the pixel
centers. The $a$ and $b$ above assume integer values in the range $[0,
  N_x-1]$ and $[0, N_y-1]$ respectively.  The discrete points on the
reciprocal plane, where $\tilde{g}$ is defined are specified by
\begin{eqnarray}
k_a = {1 \over \Delta x} \left({1 + a \over N_x} - {1 \over 2} \right) \\
k_b = {1 \over \Delta y} \left({1 + b \over N_y} - {1 \over 2} \right),
\end{eqnarray}
where $\Delta x = \Delta y$ is the pixel size.  The corresponding
inverse Fourier transform, denoted $\mathrm{IFT}[\tilde{g}],$ is
defined as
\begin{equation}
g({\bf x}) = g(x_m \hat{\imath} + y_n \hat{\jmath}) =
\sum_{a = 0}^{N_x -1}
\sum_{b = 0}^{N_y -1}
\tilde{g}(k_a \hat{\imath} + k_b \hat{\jmath})
e^{2 \pi j a m / N_x }
e^{2 \pi j b n / N_y }.
\end{equation}
According to these conventions, the discrete convolution theorem
takes on the form
\begin{equation}
\sum_l h({\bf x_l} - {\bf x_p}) g({\bf x_l}) = 
N_\mathrm{pixel} \sum_a \tilde{h}^*({\bf k_a}) \tilde{g}({\bf k_a})
\displaystyle{e^{2 \pi j {\bf k_a} \cdot {\bf x_p}}},
\label{app_conv}
\end{equation}
when $h$ and $g$ are real functions.  In Eq.~\ref{app_conv}, $l$ is an
index over all pixels of the map rather than a single dimension and
$a,$ similarly, is an index over all sampled points on the 2-d
reciprocal plane.

\bibliographystyle{hapj}
\bibliography{mm}

\end{document}